\def\im{\dot\imath}
\def\gapprox{\hbox{ \lower0.5ex\hbox{$\sim$} \kern-1.1em 
                 \raise0.5ex\hbox{$>$} }} 
\def\lapprox{\hbox{ \lower0.5ex\hbox{$\sim$} \kern-1.1em 
                 \raise0.5ex\hbox{$<$} }} 
\documentstyle[11pt, aaspp4]{article}

\begin{document}

\title{DYNAMICS OF THE KUIPER BELT}

\vskip .2truein
\author{RENU MALHOTRA}
\affil{Lunar and Planetary Institute}
\vskip .1truein
\author{MARTIN DUNCAN}
\affil{Queen's University}
\vskip .1truein
\author{and}
\vskip .1truein
\author{HAROLD LEVISON}
\affil{Southwest Research Institute}
\vskip 1truein

\begin{abstract}
Our current knowledge of the dynamical structure of the Kuiper Belt is
reviewed here.  Numerical results on long term orbital evolution
and dynamical mechanisms underlying the transport of objects
out of the Kuiper Belt are discussed. Scenarios about the origin of the
highly non-uniform orbital distribution of Kuiper Belt objects 
are described, as well as the constraints these provide on the formation 
and long term dynamical evolution of the outer Solar system.  
Possible mechanisms include an early history of orbital migration of the 
outer planets, a mass loss phase in the outer Solar system and scattering
by large planetesimals.  The origin and dynamics of the scattered component 
of the Kuiper Belt is discussed.  Inferences about the primordial mass 
distribution in the trans-Neptune region are reviewed.
Outstanding questions about Kuiper Belt dynamics are listed.
\end{abstract}

\vskip 1truein
\hrule
\begin{center}
To appear in {\it Protostars and Planets IV}\\
Preprint: January 1999
\end{center}

\vfill\eject

\noindent{\bf I.~~INTRODUCTION}

\noindent
In the middle of this century, Edgeworth~(1943) and Kuiper~(1951) independently
suggested that our planetary system is surrounded by a disk of
material left over from the formation of planets.  Both authors
considered it unlikely that the proto-planetary disk was
abruptly truncated at the orbit of Neptune.  Each also suggested that
the density in the Solar nebula was too small beyond Neptune for a major
planet to have accreted, but that this region may be inhabited by a
population of planetesimals.  Edgeworth~(1943) even suggested that bodies
from this region might occasionally migrate inward and
become visible as short-period comets. These ideas lay largely dormant
until the 1980's, when dynamical simulations  
(Fernandez~1980; Duncan, Quinn, \& Tremaine~1988; Quinn, Tremaine, \&
Duncan~1990) suggested that a disk of trans-Neptunian objects, 
now known as the {\it Kuiper belt}, was a much more likely source of the 
Jupiter-family short period comets than was the distant and isotropic
Oort comet cloud.

With the discovery of its first member in 1992 by Luu and Jewitt (1993),
the Kuiper belt was transformed from a theoretical construct to 
a {\it bona fide} component of the solar system. By now, on the order of
100 Kuiper Belt objects (KBOs) have been discovered
- a sufficiently large number that permits first-order estimates about the
mass and spatial distribution in the trans-Neptunian region.
A comparison of the observed orbital properties of these objects with
theoretical studies provides tantalizing clues to the formation and evolution
of the outer Solar system. Other chapters in this book 
describe the observed physical and orbital properties of
KBOs (Jewitt and Luu) and their collisional evolution (Farinella et al);
here we focus on the dynamical structure of the trans-Neptunian
region and the dynamical evolution of bodies in it.
The outline of this chapter is as follows:
in section II we review numerical results on long term dynamical stability
of small bodies in the outer solar system;
in section III we review our current understanding of the Kuiper Belt phase
space structure and dynamics of orbital resonances and chaotic transport
of KBOs;
section IV provides a discussion of resonance sweeping and other
mechanisms for the origin and properties of KBOs at Neptune's mean motion
resonances;
the origin and dynamics of the `scattered disk' component of the Kuiper Belt
is discussed in section V;
current ideas about the primordial Kuiper Belt are described in section VI;
we conclude in section VI with a summary of outstanding questions in
Kuiper Belt dynamics.

\bigskip
\noindent{\bf {I}{I}.~~LONG TERM ORBITAL STABILITY IN THE OUTER SOLAR SYSTEM}

\noindent{\bf A.~~Trans-Neptune region}

\noindent
We first consider the orbital stability of test particles in the 
trans-Neptunian region and the implications for the resulting structure
of the Kuiper belt several Gyrs after its presumed formation.
Torbett (1989) performed direct numerical integration of test particles 
in this region including the perturbative effects of the four giant planets,
although the latter were taken to be on fixed Keplerian orbits.
He found evidence for chaotic motion with an inverse Lyapunov
exponent (i.e. timescale for divergence of initially adjacent
orbits) on the order of Myrs for moderately eccentric,
moderately inclined orbits with perihelia between 30 and 45 AU (a
``scattered disk'').  Torbett \& Smoluchowski (1990) extended
this work and suggested that even particles with initial
eccentricities as low as 0.02 are typically on chaotic
trajectories if their semimajor axes are less than 45 AU.  Except
in a few cases, however, the authors were unable to follow the
orbits long enough to establish whether or not most chaotic
trajectories in this group led to encounters with Neptune.
Holman \& Wisdom (1993) and Levison \& Duncan (1993) showed that indeed
some objects in the
the belt were dynamically  unstable on timescales of Myr--Gyr, 
evolving onto Neptune-encountering orbits,
thereby potentially providing a source of Jupiter--family comets at the
present epoch.

Duncan, Levison \& Budd (1995) performed integrations of thousands of particles 
for up to 4 Gyr
in order to complete a dynamical survey of the trans-Neptunian region. 
The main results can be seen in Fig.~1 and include the following features: 

\noindent {\it i)}
For nearly circular, 
very low inclination particles there is a relatively stable band 
between 36 and 40 AU, with essentially complete stability beyond 42 AU. 
The lack of observed KBOs in the region between 36 and 42 AU suggests that
some mechanism besides the dynamical effects of the planets in their current 
configuration must be responsible for the orbital element distribution in 
the Kuiper belt.  This mechanism is almost certainly linked to the formation of
the outer planets.  Several possible mechanisms are described below.

\noindent {\it ii)}
For higher eccentricities (but still
very low inclinations) the region interior to 42 AU is largely unstable
except for stable bands near mean-motion resonances with Neptune
(e.g. the well-known 2:3 near 39.5 AU within which lies Pluto). 
The boundaries of the stable regions for each resonance have been
computed independently by Morbidelli et al.~(1995)
and Malhotra (1996), and are in good agreement with the results
shown in Fig.~1. The dynamics within the mean motion
resonances are discussed in detail in section III.

\noindent {\it iii)}
The dark vertical bands between 35-36 AU and 40-42 AU 
are particularly unstable regions in which the particles' eccentricities
are driven to sufficiently high values that they encounter Neptune.
These regions match very well the locations of the $\nu_7$ and $\nu_8$
secular resonances as computed analytically by Knezevic et al.~(1991)
-- i.e., the test particles precess in these regions with 
frequencies very close to two of the characteristic secular frequencies
of the planetary system.

\noindent {\it iv)}
A comparison of the observed orbits of KBOs with the phase space structure
(Figs.~1,2) shows that virtually all of the bodies interior to 42 AU are on
moderately eccentric orbits and located in mean-motion resonances,
whereas most of those beyond 42 AU appear to be in non-resonant orbits
of somewhat lower eccentricities.
In addition, there is one observed object, 1996 TL66, with semimajor axis
$\approx 80$ AU, beyond the range of Figs.~1,2.
It is thought to be a member of a third class of KBOs representing a
``scattered disk'' (see section V).  Attempts to understand the origin of
these three broad classes of KBOs will occupy much of what follows in this
chapter.

\vfill\eject

\medskip
\noindent{\bf B.~~Test particle stability between Uranus and Neptune}

\noindent
Gladman \& Duncan (1990) and Holman \& Wisdom (1993) performed long term
integrations, up to several hundred million years duration,
of the evolution of test particles on initially circular orbits in-between
the giant planets' orbits.
The majority of the test particles were
perturbed into a close approach to a planet on timescales of
0.01 -- 100 Myr, suggesting that these regions should largely 
be clear of residual planetesimals.  However, Holman (1997) has shown
that there is a narrow region, 24 -- 26 AU, lying between the orbits
of Uranus and Neptune in which roughly 1\% of minor bodies could survive
on very low eccentricity and low inclination orbits for the age of the
solar system.  He estimated that a belt of mass totaling roughly
$10^{-3}M_\oplus$ cannot be ruled out by current observational surveys.
This niche is, however, extremely fragile.  Brunini \& Melita (1998)
have shown that any one of several likely perturbations
(e.g. mutual scattering, planetary migration, and Pluto-sized perturbers)
would have largely eliminated such a primordial population.
We note also that there are similarly stable (possibly even less `fragile'),
but apparently unpopulated dynamical niches in the outer asteroid belt
(Duncan 1994) and the inner Kuiper belt (Duncan et al 1995). 

\medskip
\noindent{\bf C.~~Neptune Trojans}

\noindent
The only observational survey of which we are aware specifically designed
to search for Trojans of planets other than Jupiter covered 20 square degrees
of sky to limiting magnitude $V\!=\!22.7$ (Chen el al$.$~1997).  Although
93 Jovian Trojans were found, no Trojans of Saturn, Uranus or Neptune
were discovered.  Although this survey represents the state-of-the-art,
it lacks the sensitivity and areal coverage to reject the possibility that
Neptune holds Trojan swarms similar in magnitude to those of Jupiter
(Jewitt~1998, personal communication).
Further searches clearly need to be done.

Several numerical studies of orbital stability in Neptune's Trojan regions
have been published.  Mikkola \& Innanen~(1992) studied the behavior of
11 test particles initially near the Neptune Trojan points for 
$2 \times 10^6$ years.
Holman \& Wisdom (1993) performed a $2 \times 10^7$ year integration
of test particles initially in near-circular orbits near the Lagrange points
of all the outer planets.  And most recently, Weissman \& Levison (1997)
integrated the orbits of 70 test particles in the L4 Neptune Trojan zone
for 4 Ga. In all these studies, some Neptune Trojan orbits were found
to be stable.  Weissman \& Levison~(1997) found that stable Neptune Trojans
must have libration amplitudes $D\lapprox 60^\circ$ and proper eccentricities
$e_p\lapprox 0.05$.  It is interesting to note that this range for $D$ is 
similar to that Levison et al$.$~(1997) found for the Jupiter Trojans, but
the maximum stable $e_p$ for the Neptune regions is a factor of three smaller
than that of the Jupiter regions.
Holman \& Wisdom reported a curious asymmetric displacement of the L4 and L5
Trojan libration centers of Neptune whose cause remains unknown.

\bigskip
\noindent{\bf {I}{I}{I}.~~RESONANCE DYNAMICS AND CHAOTIC DIFFUSION}

\noindent
It is evident from the numerical analysis of test particle stability
in the trans-Neptunian region that the timescales for orbital instability
span several orders of magnitude and are very sensitive to orbital parameters.
For example, the map of stability time (i.e., time to first encounter
within a Hill sphere radius of Neptune) for initially circular orbits is
very patchy, with short dynamical lifetimes interspersed amongst very
stable regions (see Fig.~1).
Most particles that have a close approach to Neptune are removed from
the Kuiper Belt shortly thereafter by means of a quick succession of
close approaches to the giant planets. However, a small fraction evolve 
into anomalously long-lived chaotic orbits beyond Neptune that do not have
a second close approach to the planet on timescales comparable to the age
of the Solar system (see section V).
The nature and origin of the long-timescale instabilities (which are most
relevant for understanding the origin of short period comets from the Kuiper
Belt) is not well understood at present.

In general, we understand that the mean motion resonances of Neptune
form a `skeleton' of the phase space (Fig.~2),
with the perturbations of the other giant planets,
including secular resonances, forming a web of superstructure on that skeleton.
The phase space in the neighborhood of Neptune's 3:2 mean motion resonance
is the best studied, following the discovery of Pluto and its myriad of
resonances (cf., Malhotra \& Williams, 1997).
Fig.~3 shows the dynamical features that have been identified at the 3:2
resonance, in the $(a,e)$ plane.
The boundary of this resonance, determined by means of a semi-numerical
analysis of the averaged perturbation potential of Neptune, is shown
(dark solid lines), as well as the loci of the apsidal $\nu_8$ and nodal
$\nu_{18}$ secular resonances and the Kozai resonance in this neighborhood
(Morbidelli, 1997).
The stable libration zone, determined from an inspection of many surfaces of
section of the planar restricted 3-body model of the Sun-Neptune-Plutino 
(Malhotra, 1996), is indicated by the blue shaded region.
The stable resonance libration boundary is significantly different from
the formal perturbation theory result because averaging is not a good
approximation in the vicinity of the resonance separatrix:
the separatrix broadens into a chaotic zone owing to the interaction with
neighboring mean motion resonances. The width of the chaotic separatrix
increases with eccentricity, eventually merging with the chaotic separatrices
of neighboring mean motion resonances.
The $\nu_8$ secular resonance is mostly embedded in the chaotic zone,
while the $\nu_{18}$ occurs at large libration amplitudes close to the
chaotic separatrix; the Kozai resonance occurs at large libration amplitude
for low eccentricity orbits, and at smaller libration amplitude for
eccentricities near 0.2--0.25.

Fig.~4 shows a ``map'' of the diffusion speed of test particles determined
by numerical integrations of up to 4 billion years by Morbidelli (1997).  
It is clear that instability timescales in the vicinity of the 3:2 resonance
range from less than a million years to longer than the age of the Solar system.
In general, within the resonance, higher eccentricity orbits are less stable
than lower eccentricities; the stability times are longest deep in the
resonance and shorter near the boundaries. 
We note that the uncolored regions exterior and adjacent to the colored zones
at eccentricities below $\sim0.15$ are stable for the age of the Solar system,
but those above $\sim0.15$ are actually chaotic on timescales shorter than the
shortest indicated in the colored zones.

The short stability timescales in the most unstable zones are due to dynamical
chaos generated by the interaction with neighboring mean motion resonances;
this can be directly visualized in surfaces of section of the planar restricted
three body model (Malhotra, 1996).
Test particles in these zones suffer large chaotic changes in semimajor axis
and eccentricity on short timescales, ${\cal O}(10^{5-7})$ yr, and are not
protected from close encounters with Neptune.
In a small zone near semimajor axis 39.5 AU,
initially circular low-inclination orbits are excited to high eccentricity
{\it and} high inclination on timescales of ${\cal O}(10^{7})$ yr
(Holman \& Wisdom (1993), Levison \& Stern (1997));
this instability is possibly due to overlapping secondary resonances
(Morbidelli, 1997).
The finite diffusion timescale, comparable to but shorter than the age of
the Solar system, in a large area (green zone) inside the resonance is not
understood at all; possibly higher order secondary resonances are the
underlying cause.
The numerical evolution of test particles in this region follows initially
a slow diffusion in semimajor axis with nearly constant mean eccentricity
and inclination, until the orbit eventually reaches a strongly chaotic zone
(Morbidelli, 1997).
The diffusion timescales in this region are comparable to or only slightly
less than the age of the Solar system, so that it is likely an active source
region for short period comets via purely dynamical instabilities.

The dynamical structure in the vicinity of other mean motion resonances 
has not been obtained in as much detail as that of the 3:2. We expect
differences in details -- different profile of the libration zone,
differing secular resonance effects, etc. --
but generally similar qualitative features.
We also note that since the orbital evolution obtained in
non-dissipative models of Kuiper Belt dynamics is time-reversible,
the transport of particles from strongly chaotic regions to weakly chaotic
regions is also allowed.  In the most general terms, this is the likely
explanation for the putative Scattered Disk (section V).

\bigskip
\noindent{\bf {I}{V}.~~RESONANT KUIPER BELT OBJECTS}

\noindent
The origin of the great abundance of resonant KBOs and, equally importantly,
their high orbital eccentricities, is an interesting question
whose understanding may lead to significant advances in our understanding of
the early history of the Solar system.
In this section, we discuss current ideas pertinent to this class of KBOs.

\medskip
\noindent{\bf A.~~Planet migration and resonance sweeping}

\noindent
An outward orbital migration of Neptune in early solar system history
provides an efficient mechanism for sweeping up large numbers of 
trans-Neptunian objects into Neptune's mean motion resonances.
We describe this scenario in some detail here; its importance stems from
the linkage it provides between the detailed orbital distribution in the
Kuiper Belt and the early orbital migration history of the outer planets.
This theory, which was originally proposed for the origin of Pluto's orbit
(Malhotra, 1993), supposes that Pluto formed in a common low-eccentricity,
low-inclination orbit beyond the (initial) orbit of Neptune.
It was captured into the 3:2 resonance with Neptune and had its eccentricity
excited to its current Neptune-crossing value as Neptune's orbit expanded
outward due to angular momentum exchange with residual planetesimals.
The theory predicts that resonance capture and eccentricity excitation would
be a common fate of a large fraction of trans-Neptunian objects
(Malhotra, 1995).

The process of orbital migration of Neptune (and, by self-consistency,
migration of the other giant planets) invoked in this theory is as follows.
Consider the late stages of planet formation when the outer
Solar system had reached a configuration close to its present state,
namely four giant planets in well separated near-circular, co-planar
orbits, the nebular gas had already dispersed, the planets had accreted
most of their mass, but there remained a residual population of icy
planetesimals and possibly larger planetoids.
The subsequent evolution consisted of the gravitational scattering and
accretion of these small bodies.
Circumstantial evidence for this exists in the obliquities of the planets
(Lissauer \& Safronov, 1991; Dones \& Tremaine, 1993).
Much of the Oort Cloud -- the putative source of long period comets -- would
have formed during this stage by the scattering of planetesimals to wide orbits
by the giant planets and the subsequent action of galactic tidal perturbations
and perturbations from passing stars and giant molecular clouds
(e.g.~Fernandez, 1985; Duncan et al, 1987).
During this era, the back reaction of planetesimal scattering on the planets
could have caused significant changes in their orbital energy and angular
momentum.  Overall, there was a net loss of energy and angular momentum from
the planetary orbits, but the loss was not extracted uniformly from the four
giant planets.  Jupiter, by far the most massive of the planets,
likely provided all of the lost energy and angular momentum, and more;
Saturn, Uranus and Neptune actually gained orbital energy and angular momentum
and their orbits expanded, while Jupiter's orbit decayed sufficiently
to balance the books.
This was first pointed out by Fernandez \& Ip (1984) who noticed it in
numerical simulations of the late stages of accretion of planetesimals by
the proto-giant planets.  

\medskip
\noindent {\it Migration of the Jovian planets}

\noindent
The reasons for the rather non-intuitive result can be understood from
the following heuristic picture of the clearing of a planetesimal swarm
from the vicinity of Neptune.
Suppose that the mean specific angular momentum of the swarm is initially
equal to that of Neptune.  A small fraction of planetesimals is accreted as
a result of physical collisions. Of the remaining, there are approximately
equal numbers of inward and outward scatterings.
To first order, these cause no net change in Neptune's orbit.  However,
the subsequent fate of the inward and outward scattered planetesimals
is not symmetrical. Most of the inwardly scattered objects enter the zones
of influence of Uranus, Saturn and Jupiter.
Of those objects scattered outward, some are eventually lifted into wide,
Oort Cloud orbits while most return to be rescattered;
a fraction of the latter is again rescattered inwards where the inner
Jovian planets, particularly Jupiter, control the dynamics.
The massive Jupiter is very effective in causing a systematic loss of
planetesimal mass by ejection into Solar system escape orbits.
As Jupiter preferentially removes the inward scattered planetesimals from
Neptune's influence, the planetesimal population encountering Neptune at
later times is increasingly biased towards objects with specific angular
momentum and energy larger than Neptune's.  Encounters with this planetesimal
population produce effectively a negative drag on Neptune which increases its
orbital radius.  Uranus and Saturn, also being much less massive
than and exterior to Jupiter, experience a similar orbital migration, but
smaller in magnitude than Neptune.
Thus Jupiter is, in effect, the source of the angular momentum and energy
needed for the orbital migration of the outer giant planets, as well as for
the planetesimal mass loss. However, owing to its large mass, its orbital
radius decreases by only a small amount.

The magnitude and timescale of the radial migration of the Jovian planets due
to their interactions with residual planetesimals is difficult to determine
without a full-scale N-body model.
The work of Fernandez \& Ip (1984) is suggestive but not conclusive
due to several limitations of their numerical model which produced highly
stochastic results: they modeled a small number of planetesimals, $\sim2000$,
and the masses of individual planetesimals were in the rather exaggerated
range of 0.02--0.3 $M_\oplus$; and, perhaps most significantly,
they neglected long range gravitational forces.
Current studies attempt to overcome these limitations by using the
faster computers now available and more sophisticated integration algorithms
(Hahn \& Malhotra, 1998).  Still, fully self-consistent high fidelity models
remain a distant goal at this time.
Remarkably, an estimate for the magnitude and timescale of Neptune's outward
migration is possible from an analysis of the orbital evolution of KBOs
captured in Neptune's mean motion resonances.

\medskip
\noindent{\it Resonance sweeping}

\noindent
Capture into resonance is a delicate process, difficult to analyze
mathematically.  Under certain simplifying assumptions, Malhotra (1993, 1995)
showed that resonance capture is very efficient for adiabatic orbital
evolution of KBOs whose initial orbital eccentricities were smaller than
$\sim0.05$.
Resonance capture leads to an excitation of orbital eccentricity whose
magnitude is related to the magnitude of orbital migration:
$$ \Delta e^2 \simeq {k\over j+k} \ln {a_f\over a_i}
= {k\over j+k} \ln {a_N\over a_{N,i}}, \eqno(1)$$
where $a_i$ and $a_f$ are the initial and current semimajor axes of a KBO,
$a_N$ is Neptune's current semimajor axis and $a_{N,i}$ is its value
in the past at the time of resonance capture; $j$ and $k$ are positive
integers defining a $j:j+k$ mean motion resonance.
From this equation and the observed eccentricities of Pluto and the Plutinos
(see Jewitt \& Luu, this vol.),
it follows that Neptune's orbit has expanded by $\sim\!9$ AU.

Numerical simulations of resonance sweeping of the Kuiper Belt have been
carried out assuming adiabatic giant planet migration of specified magnitude
and timescale.  The orbital distribution of Kuiper Belt objects obtained in
one such simulation is shown in Fig.~5.
The main conclusions from such simulations are that
(i) few KBOs remain in circular orbits of semimajor axis $a\lapprox 39$ AU which
marks the location of the 3:2 Neptune resonance;
(ii) most KBOs in the region up to $a=50$AU are locked in mean motion
resonances; the 3:2 and 2:1 are the dominant resonances, but the 4:3 and 5:3 
also capture noticeable numbers of KBOs;
(iii) there is a significant paucity of low-eccentricity orbits in the 3:2
resonance; 
(iv) the maximum eccentricities in the resonances are in excess of
Neptune-crossing values;
(v) inclination excitation is not as efficient as eccentricity excitation:
only a small fraction of resonant KBOs acquire inclinations in excess of
$10^\circ$.
Not shown in Fig.~5 are other dynamical features such as the resonance
libration amplitude and the argument-of-perihelion behavior (libration, as
for Pluto, or circulation), which are also reflective of the planet
migration/resonance sweeping process.

More detailed discussion of numerical results on resonance sweeping is given
in Malhotra (1995,1997,1998a,1999) and Holman (1995).
Two additional points are worthy of note here.
One is that, owing to the longer dynamical timescales associated with vertical
resonances, the magnitude of inclination excitation of KBOs is sensitive to both
the magnitude and the timescale of planetary migration; it is estimated that a
timescale on the order of $(1-3)\times10^7$ yr would account for Pluto's inclination
(Malhotra, 1998a).
The second is that the total mass of residual planetesimals required for a
$\sim9$ AU migration of Neptune is estimated to be $\sim\!50M_\oplus$ (Malhotra, 1997);
this estimate is supported by recent numerical simulations of planet migration
(Hahn \& Malhotra, 1998).

An outstanding issue is the apparent paucity of KBOs at Neptune's 2:1 mean
motion resonance in the observed sample (Fig.~2)\footnote{In December 1998,
while this article was in the review process, the Minor Planet Center
reported new observations yielding revised orbits for 1997 SZ10 and 1996 TR66,
identifying these as the first two KBOs in the 2:1 resonance with Neptune
(Marsden, 1998).}, whereas the planet migration/resonance sweeping theory
predicts comparable populations in the 3:2 and 2:1 resonances (Fig.~5).
Possible explanations are:
(i) observational incompleteness (cf., Gladman et al, 1998), or
(ii) significant leakage out of the 2:1 resonance on billion year timescales
by means of weak instabilities or perturbations by larger members of the
Kuiper Belt (Malhotra, 1999), or
(iii) the planet migration/resonance sweeping did not occur as postulated.
However, the success of the resonance sweeping mechanism in explaining the 
orbital eccentricity distribution of Plutinos --
and the difficulty of explaining it by other means --
argues strongly in its favor.
Further work is needed to refine the relationship between the orbital
element distributions and the detailed characteristics of the planet
migration process, including the overall efficiency of resonance capture
and retention.

If such planet migration and resonance sweeping occurred, then it follows
that the KBOs presently resident in Neptune's mean motion resonances 
formed closer to the Sun than their current semimajor axes would suggest.
If there were a significant compositional gradient in the primordial
trans-Neptunian planetesimal disk, it may be preserved in a rather subtle
manner in the present orbital distribution.
Because each resonant KBO retains memory of its initial orbital radius in its
final orbital eccentricity (Eq.~1), there would exist a compositional
gradient with orbital eccentricity within each resonance;
non-resonant KBOs in near-circular low-inclination orbits between 30 AU and
50 AU most likely formed at their present locations and would reflect the
primordial conditions at those locations.
However, if there were significant orbital mixing by processes other than
resonance sweeping, these systematics would be diluted or erased.

\medskip
\noindent{\bf B.~~Secular resonance sweeping}

\noindent
The combined, orbit-averaged perturbations of the planets on each other
cause a slow precession of the direction of perihelion and of the pole
of the orbit plane of each planet.
Of particular importance to the long term dynamics of the Kuiper Belt
are the perihelion and orbit pole precession of Neptune, both of
which have periods of about 2 Myr in the present planetary configuration.
The perihelion direction and orbit pole orientation of the orbits of
Kuiper Belt objects also precess slowly, at rates that depend upon their
orbital parameters.  For certain ranges of orbital parameters,
the perihelion precession rate matches that of Neptune; this condition
is termed the $\nu_8$ {\it secular resonance}.
Similarly, the 1:1 commensurability of the rate of precession of the
orbit pole with that of Neptune's orbit pole is called $\nu_{18}$
{\it secular resonance}. 
(See Malhotra (1998b) for an analytical model of secular resonance.)
The $\nu_8$ and $\nu_{18}$ secular resonances occur at several
regions in the Kuiper Belt where they cause strong perturbations
of the orbital eccentricity and orbital inclination, respectively
(cf.~Fig.~1,3).

The secular effects are sensitive to the mass distribution in the
planetary system (see Ward 1981, and references therein).
Levison et al.~(1999) have noted that a primordial massive
trans-Neptunian disk would have significantly altered
the locations of the $\nu_8$ and $\nu_{18}$ secular resonances.
From a suite of numerical simulations,
they estimate that a $\sim10M_\oplus$ primordial disk between 30 AU and 50 AU
would have the $\nu_8$ secular resonance near $a\lapprox36$ AU,
and that it would have moved outward to its current location near 42 AU
as the disk mass declined.
Such sweeping by the $\nu_8$ secular resonance excites the orbital
eccentricities of KBOs sufficiently to cause them to encounter Neptune
and be removed from the Kuiper Belt.
Only objects fortuitously trapped in Neptune's mean motion resonances
remain stable.  
The simulations also find that the $\nu_{18}$ secular resonance 
sweeps {\it inward} from well beyond its current location as the Kuiper
Belt mass declines, thereby moderately increasing the inclinations
(up to $\sim 15$ degrees) of KBOs beyond 42 AU.
However, this mechanism does not produce orbital eccentricities in Neptune's
2:3 mean motion resonance as large as those observed.
Furthermore, damping of the eccentricity and inclination by density waves
is to be expected in a massive primordial disk (Ward \& Hahn, 1998),
but has not yet been included in the simulations.
We conclude that the sweeping of secular resonances has probably played
some role in the excitation of the Kuiper belt, but its quantitative effects
remain to be determined.

\medskip
\noindent{\bf C.~~Stirring by Large Neptune-scattered Planetesimals}

\noindent
A third mechanism to explain the observed orbital properties of KBOs
is to invoke the orbital excitation produced by close encounters with
large Neptune-scattered planetesimals on their way out of the solar
system or to the Oort cloud.  The observed excitation in the Kuiper
belt requires the prior existence of planetesimals
with masses $\sim 1$ Earth mass according to Morbidelli and Valsecchi (1997). 
There is circumstantial evidence (e.g. the obliquity of Uranus' spin axis)
that a population of objects this massive might have
formed in the region between Uranus and Neptune and are now
gone (Safronov 1966, Stern 1991).  Many of these objects must have 
spent some time orbiting through some parts of the Kuiper belt.
A much more massive initial belt might then have been sculpted to its
presently observed structure because of the injection of most of the
small bodies into dynamically unstable regions in the inner belt and
the enhanced role of mutually catastrophic collisions among small
planetesimals in the outer belt.  In this picture, then, the observed
KBOs interior to $\sim 42$ AU are the lucky ones ending up in the
small fraction of phase space ($\sim$ few percent --see Figure 2)
protected from close encounters with Neptune by mean-motion
resonances.  This mechanism is similar to
one involving large Jupiter-scattered planetesimals proposed earlier
to explain the excitation of the asteroid belt
(Ip 1987; Wetherill 1989).

Petit et al. (1998) have combined 3-body integrations and semi-analytic
estimates of scattering to model the effects of large planetesimals
in the Kuiper belt. They argue that the best reconstruction of the observed
dynamical excitation of the Kuiper belt requires  
the earlier existence of two large bodies. The first is a body of half an 
Earth mass on an orbit of large eccentricity with a dynamical lifetime of 
several times $10^8$ yr. The second is a body of 1 Earth mass, 
which evolves for $\sim 25$ Myr on a low eccentricity orbit spanning the 
30-40 AU range.

An attractive feature of this mechanism is that it yields an overall
mass depletion in the inner Kuiper Belt and accounts for the fact
that the outer Kuiper belt ($a > 42$ AU) is moderately excited. 
It does not, however, appear to explain the lack of low-eccentricity
objects in Neptune's 2:3.  In addition, the models performed to
date require a specific set of very large objects, for which there is no
direct evidence, to be at the right place for the right length of time. 
The presumed eventual removal of these objects by means of a final close
encounter with Neptune would perturb Neptune's orbit significantly,
and also jeopardize the stability of resonant KBOs;
this is in conflict with the observed evidence.

\bigskip
\noindent{\bf {V}.~~THE SCATTERED DISK}

\noindent
As noted previously, the current renaissance in
Kuiper belt research was prompted by the suggestion that the
Jupiter-family comets originated there (Fern\'andez~1980; Duncan,
Quinn, \& Tremaine~1988).  Thus, as part of the research intended to
understand the origin of these comets, a significant amount of
effort has gone into understanding the dynamical behavior of objects
that are on orbits that can encounter Neptune (Duncan, Quinn, \&
Tremaine~1988; Quinn, Tremaine, \& Duncan~1990; Levison \& Duncan
1997).  It is somewhat ironic, therefore, that these studies have led
to the realization that a structure known as the {\it Scattered Comet
Disk}, rather than the Kuiper belt, could be the dominant source of the
Jupiter-family comets.

For our purposes, scattered disk objects are distinct from Kuiper
belt objects in that they evolved out of their primordial orbits
beyond Uranus early in the history of the solar system.  These objects
were then dynamically scattered by Neptune into orbits with perihelion
distances near Neptune, but semi-major axes in the Kuiper belt (Duncan
\& Levison~1997).  Finally, some process, usually interactions with
Neptune's mean motion resonances, raised their perihelion distances
thereby effectively storing the objects for the age of the solar
system.  Scattered disk objects (hereafter SDOs) occupy the same
physical space as KBOs, but can be distinguished from KBOs by their
orbital elements.  In particular, as we describe in more detail below,
SDOs tend to be on much more eccentric orbits than KBOs.
  
Of the 60 or so Kuiper belt objects thus far cataloged, only one,
1996~TL$_{66}$, is an obvious SDO.  1996~TL$_{66}$ was discovered in
October, 1996 by Jane Luu and colleagues (Luu et~al$.$~1997) and is
estimated to have a semi-major axis of 85 AU, a perihelion of 35 AU,
an eccentricity of 0.59, and an inclination of $24^\circ$.  Such a
high eccentricity orbit most likely resulted from gravitational
scattering by a giant planet, in this case Neptune.

The idea of the existence of a scattered comet disk dates to
Fern\'andez \& Ip~(1989). Their numerical simulations indicated that
some objects on eccentric orbits with perihelia inside the orbit of
Neptune could remain on such orbits for billions of years and hence
might be present today. However, their simulations were based on an
algorithm which incorporates only the effects of close
gravitational encounters (\"Opik, 1951),
and hence severely overestimates the
dynamical lifetimes of bodies such as those in their putative disk
(Dones et al, 1998). As a result, the dynamics of their scattered
disk bears little resemblance to the structure found in more recent
direct integrations to be discussed below.

The only investigation of the scattered disk which uses modern direct
numerical integrations is the one by Duncan \& Levison~(1997,
hereafter DL97).  DL97 was a followup to Levison \& Duncan~(1997),
which was an investigation of the behavior of 2200 small, massless
objects that initially were encountering Neptune.  The latter's focus
was to model the evolution of these objects down to Jupiter-family
comets and followed the system for only 1 billion years.  DL97
extended these integrations to 4 billion years.  Most objects that
encounter Neptune have short dynamical lifetimes.  Usually, they are
either 1) ejected from the solar system, 2) hit the Sun or a planet,
or 3) are placed in the Oort cloud, in less than $\sim 5
\times10^{7}$ years.  It was found, however, that 1\% of the particles
remained in orbits beyond Neptune after 4 billion years.  So, if at
early times there was a significant amount of material from the region
between Uranus and Neptune or the inner Kuiper belt that evolved onto
Neptune-crossing orbits, then there could be a significant amount of
this material remaining today.  What is meant by `significant' is the
main question when it comes to the current importance of the scattered
comet disk.

DL97 found that some of the long-lived objects were scattered to very
long-period orbits where encounters with Neptune became
infrequent. However, at any given time, the majority of them were
interior to 100 AU.  Their longevity is due in large part to their
being temporarily trapped in or near mean-motion resonances with
Neptune.  The `stickiness' of the mean motion resonances, which was
mentioned by Holman \& Wisdom~(1993), leads to an overall
distribution of semi-major axes for the particles that is peaked near
the locations of many of the mean motion resonances with Neptune.
Occasionally, the longevity is enhanced by the presence of the Kozai
resonance.
In all long-lived cases, particles had their perihelion distances
increased so that close encounters with Neptune no longer
occurred. Frequently, these increases in perihelion distance were
associated with trapping in a mean motion resonance, although in many
cases it has not yet been possible to identify the exact process that
was involved. On occasion, the perihelion distance can become large,
but 81\% of scattered disk objects have perihelia between 32 and $36\,
AU$.

Fig.~6 shows the dynamical behavior of a typical particle.  This
object initially underwent a random walk in semi-major axis due to
encounters with Neptune.  At about $7 \times10^7$ years it was temporarily
trapped in Neptune's 3:13 mean motion resonance for about $5 \times10^7$
years.  It then performed a random walk in semi-major axis until about
$3 \times10^8$ years, when it was trapped in the 4:7 mean motion resonance,
where it remained for $3.4 \times10^9$ years.  Notice the increase in the
perihelion distance near the time of capture.  While trapped in this
resonance, the particle's eccentricity became as small as
$0.04$. After leaving the 4:7, it was trapped temporarily in Neptune's
3:5 mean motion resonance for $\sim 5 \times10^8$ yr and then went through
a random walk in semi-major axis for the remainder of the simulation.

DL97 estimated an upper limit on the number of possible SDOs by
assuming that they are the sole source of the Jupiter-family comets.
DL97 computed the simulated distribution of comets throughout the
solar system at the current epoch (averaged over the last billion
years for better statistical accuracy).  They found that the ratio of
scattered disk objects to visible Jupiter-family comets\footnote{We
define a `visible' Jupiter-family comet as one with a perihelion
distance less than $2.5\, AU$.} is $1.3 \times10^6$.  Since there are
currently estimated to be 500 visible JFCs (Levison \& Duncan~1997),
there are $\sim 6 \times 10^8$ comets in the scattered disk if it is
the sole source of the JFCs.
It is quite possible that the scattered disk could contain this much material.
Figure~7 shows the spatial distribution for this model.

To review the above findings, $\sim 1\%$ of the objects in
the scattered disk remain after 4 billion years, and that $6 \times
10^8$ comets are currently required to supply all of the 
Jupiter-family comets.  Thus, a scattered comet disk requires an initial
population of only $6 \times 10^{10}$ comets (or $\sim 0.4 M_\oplus$,
Weissman~1990) on Neptune-encountering orbits.  Since planet formation
is unlikely to have been 100\% efficient, the original disk could have
resulted from the scattering of even a small fraction of the tens of
Earth masses of cometary material that must have populated the outer
solar system in order to have formed Uranus and Neptune.

\bigskip
\noindent{\bf {V}{I}.~~THE PRIMORDIAL KUIPER BELT}

\noindent
As with many scientific endeavors, the discovery of new
information tends to raise more questions than it answers.  Such is
the case with the Kuiper belt.  Even the original argument that
suggested the Kuiper belt is in doubt.  Edgeworth's (1949) and
Kuiper's (1951) speculations were based on the idea that it
seemed unlikely that the disk of planetesimals that formed the planets
would have abruptly ended at the current location of the outermost
known planet, Neptune.  An extrapolation into the Kuiper belt of the current
surface density of non-volatile material in the outer planetary region
implies that there should be about $30 M_\oplus$ of material there.
However, the recent KBO observations indicate only a few $\times
\sim 0.1 M_\oplus$ between 30 and $50\, AU$.
Were Kuiper and Edgeworth wrong?  Is there a sharp outer edge to
the planetary system?  
One line of theoretical arguments suggests that the answer may be no to
both questions; we discuss these below, but we note at the end some
contrary arguments as well.

Over the last few years, several points have been made supporting the
idea of a massive primordial Kuiper belt; see Farinella et al.~(this volume).
Stern (1995) and Davis \& Farinella (1997) have argued that the
inner part of the Kuiper belt ($\lapprox 50\, AU$) is currently
eroding away due to collisions, and therefore must have been more
massive in the past.  Stern (1995) also argued that the current surface
density in the Kuiper Belt is too low to grow bodies larger than about
$30\, km$ in radius by means of two-body collisional accretion over the
age of the Solar system.  The observed $\sim 100$ km size KBOs could
have grown in a more massive Kuiper belt, of at least several Earth masses, 
if the mean eccentricities of the accreting objects were small,
on the order of a few times $10^{-3}$ or perhaps as large as $10^{-2}$
(Stern~1996; Stern \& Colwell~1997a; Kenyon \& Luu~1998).
Models of the collisional evolution of a massive Kuiper belt suggest that
90\% of the mass inside of $\sim 50\, AU$ could have been been lost due to
collisions over the age of the solar system (Stern \& Colwell~1997b;
Davis \& Farinella~1998).  Thus, it is possible for a massive primordial
Kuiper belt to grind itself down to the levels that we see today.

With these arguments, it is possible to build a strawman model of the
Kuiper belt, which is depicted in Fig.~8.  Following Stern~(1996),
there are three distinct zones in the Kuiper belt.  
Region~A is a zone where the gravitational perturbations of the outer
planets have played an important role, tending to pump up the eccentricities
of objects.  About half of the objects in this region could have been
dynamically removed from the Kuiper belt (Duncan et al. 1995).
The remaining objects have eccentricities that are large enough that
accretion has ceased and erosion is dominant.
Thus, we expect that a significant fraction of the mass in region~A
has been removed by collisions.
The dotted curve shows an estimate of the initial surface density of solids,
extrapolated from that of the outer planets, and the solid curve (marked
DLB95) shows an estimate of the current Kuiper belt surface density from
Duncan et al.~(1995, reproduced from their Fig.~7).
Region~B in Fig.~8 is a zone where collisions are important
but the perturbations of the planets are not.
The orbital eccentricities of objects in this region will
most likely remain small enough that collisions lead to accretion of
bodies, rather than erosion.  Large objects could have formed here
(see Stern \& Colwell~1997a, hereafter SC97a), but the surface density
may not have changed very much in this region over the age of the
solar system.  Observational constraints based on COBE/DIRBE data put
the transition between Regions~A and B beyond $70\, AU$
(Teplitz et al.~1999).  The outer boundary of Region~B is very uncertain, but
will likely be beyond $100\, AU$ (Stern, personal communication).
Region~C is a zone where collision rates are low enough that the
surface density of the Kuiper belt and the size distribution of
objects in it have remained virtually unchanged over the history of
the solar system. 

If the above model is correct, then the Kuiper belt that we have so
far observed may be a low density region that lies inside a much more
massive outer disk.  In other words, we may now be seeing a `Kuiper Gap'
(Stern \& Colwell~1997b).

The idea of an increase in the surface density of the Kuiper belt
beyond $50\, AU$ may explain a puzzling feature of the dynamics of the
planets: Neptune's small eccentricity.
Ward \& Hahn~(1998, hereafter WH98) have shown that if the Kuiper belt
smoothly extends to a couple of hundred $AU$ and if the eccentricities
in the disk are smaller than $\sim 0.05$ then Neptune can excite
apsidal waves in the disk that will carry away angular momentum from
Neptune's orbit and damp that planet's eccentricity.
They estimate an upper limit of $\sim 2 M_\oplus$ on the mass of material
between 50 and $75\, AU$.  This upper limit is marked WH98 in Fig.~8.

We note that WH98 estimate more mass than a simple extension
of the Kuiper belt's {\it observed} surface density would imply
(i$.$e$.$ we are seeing a Kuiper gap).
However, WH98's upper limit on the surface density is much less than
that required by the accretion models to grow the observed KBOs.
There are three possible resolutions to this problem.
$\im)$ It could be a matter of timing.  As SC97a describe, the KBOs
must have been formed before Neptune in order for their eccentricities
to be small enough for accretion.  Conversely, WH98's upper limit only applies
after Uranus and Neptune have cleared away any local objects that could
excite their eccentricities.  Perhaps the surface density of the Kuiper
belt decreased significantly between these two times.
For example, it could have decreased due to Neptune and Uranus injecting
massive objects into the 50-$100\, AU$ region which stirred up the disk
and greatly increased the collisional erosion (Morbidelli \& Valsecchi~1997).
$\im\im)$ The uncertainties in the models could be larger than the
discrepancy between them.  SC97a make assumptions about the velocity
evolution of their objects and the physics of the collisions.  WH98
make assumptions about the shape of the surface density distribution
and eccentricities.  Perhaps when more details models are constructed,
the discrepancy will vanish. 
$\im\im\im)$ There really was an edge to the disk from which the planets
formed at about $50\, AU$.  This would explain why no classical KBOs have
been discovered beyond this point.
Although the lack of discoveries may be due to observational incompleteness
(Gladman et al$.$~1998), Monte Carlo simulations of the detection statistics
of observational surveys are consistent with a Kuiper Belt edge at 50 AU
(Jewitt et al., 1998).  Clearly more work needs to be done on this topic.

\bigskip
\noindent{\bf {V}{I}{I}.~~CONCLUDING REMARKS}

\noindent
Studies of Kuiper Belt dynamics offer the exciting prospect of deriving
constraints on dynamical processes in the late stages of planet formation
which have hitherto been considered beyond observational constraint.
Many questions have been raised by the intercomparisons between observations
of the Kuiper Belt and theoretical studies of its dynamics;
these outstanding issues are listed below.

\noindent 1.
It is likely that the Kuiper Belt defines an outer boundary condition for
the primordial planetesimal disk, and by extension, for the primordial
Solar Nebula.  What new constraints does it provide on the Solar Nebula,
its spatial extent and surface density, and on the timing and manner of
formation of the outer planets Uranus and Neptune?
How does the Kuiper Belt fit in with observed dusty disks around other
sun-like stars, such as Beta Pictoris?

\noindent 2.
What is the spatial extent of the Kuiper Belt -- its radial and inclination
distribution?  What are the relative populations in the Scattered Disk
and the Classical Kuiper Belt?
What mechanisms have given rise to the large eccentricities and inclinations
in the trans-Neptunian region?

\noindent 3.
What are the relative proportions of the resonant and non-resonant KBOs,
their eccentricity, inclination and libration amplitude distributions?
These provide constraints on the orbital migration history of the outer planets.

\noindent 4.
The phase space structure in the vicinity of Neptune's mean motion resonances
is reasonably well understood only for the 3:2 resonance.  Similar studies
of the other resonances are warranted.

\noindent 5.
Given the apparent highly nonuniform orbital distribution of KBOs, precisely 
what is the source region of short period comets and Centaurs?
What is the nature of the long term instabilities that provide dynamical
transport routes from the Kuiper Belt to the short period comet
and Centaur population?

\noindent 6.
Was the primordial Kuiper Belt much more massive than at present?
What, if any, were the mass loss mechanisms
(collisional grinding, dynamical stirring by large KBOs or lost planets)?

\noindent 7.
Is there a significant population of Neptune Trojans?
Or a belt between Uranus and Neptune?
What can we learn from its presence/absence about the dynamical history
of Neptune?

\noindent 8.
What is the distribution of spins of KBOs?
It may help constrain their collisional evolution.

\noindent 9.
What is the frequency of binaries in the Kuiper Belt?
(How unique is the Pluto-Charon binary?)  

\noindent 10.
What is the relationship between the Kuiper Belt and the Oort Cloud?
How does the mass distribution in the Kuiper Belt relate to the formation
of the Oort Cloud?
Is there a continuum of small bodies spanning the two regions?

\bigskip
\noindent{\bf {V}{I}{I}{I}.~~ACKNOWLEDGEMENTS}

\noindent
The authors acknowledge support from NASA's Planetary Geosciences and
Origins of Solar Systems Research Programs.
MJD is also grateful for the continuing financial support of the Natural
Science and Engineering Research Council of Canada.
HL thanks P.~Weissman and S.A~Stern for discussions.
Part of the research reported here was done while RM was a
Staff Scientist at the Lunar and Planetary Institute
which is operated by the Universities Space Research Association
under contract no.~NASW-4574 with the National Aeronautics and Space
Administration.
This paper is Lunar and Planetary Institute Contribution no.~959.

\vfill\eject

\vfill\eject

\vskip .5in
\centerline{\bf FIGURE CAPTIONS}
\vskip .25in

\figcaption{
Dynamical lifetime before first close encounter with Neptune for test particles
with a range of semi-major axes and eccentricities, and with initial
inclinations of one degree (based on Duncan et al 1995).
Each particle is shown as a narrow vertical strip, centered on the particle's
initial eccentricity and semimajor axis. The lightest colored (yellow) strips
represent particles that survived the length of the integration 
(4 billion years). Dark regions are particularly unstable.
The dots indicate the orbits of Kuiper belt objects with reasonably
well-determined orbits in January, 1999.
(Orbits with inclinations less than 10 degrees are shown
in red, those with higher inclinations are displayed in green.) 
The locations of low order mean motion resonances with Neptune 
and two curves of constant perihelion distance, $q$ are shown.}

\bigskip

\figcaption{
The locations and widths in the $(a,e)$ plane of Neptune's low order
mean motion resonances in the Kuiper Belt.  Orbits above the dotted
line are Neptune-crossing; the hatched zone on the left indicates
the chaotic zone of first order resonance overlap.
The dots indicate the orbits of KBOs with reasonably well determined
orbits in January 1999.  (Adapted from Malhotra (1995).)}

\bigskip

\figcaption{
The major dynamical features in the vicinity of Neptune's 3:2 mean motion
resonance.  The locations of the resonance separatrix (dark solid lines),
two secular resonances, apsidal $\nu_8$ and nodal $\nu_{18}$, and the
Kozai resonance were obtained by a semi-numerical analysis of the averaged
perturbation potential of Neptune and the other giant planets.
The blue shaded region is the stable resonance libration zone in the
unaveraged potential of Neptune. (Adapted from Morbidelli (1997)
and Malhotra (1995).)}

\bigskip

\figcaption{
A map of the dynamical diffusion speed of the semimajor axis of test
particles in Neptune's 3:2 mean motion resonance.
The test particles in this numerical study had initial inclinations less
than 5 degrees.  The color scale is indicated on the bottom.
(From Morbidelli (1997).)}

\bigskip

\figcaption{
The distribution of orbital elements of trans-Neptunian test particles
after resonance sweeping, as obtained from a 200 Myr numerical simulation
in which the Jovian planets' semimajor axes evolve according to
$a(t)=a_f-\Delta a\exp(-t/\tau)$, with timescale $\tau=4$Myr and
$\Delta a=\{-0.2,0.8,3.0,7.0\}$ AU for Jupiter, Saturn, Uranus and 
Neptune, respectively (Malhotra, 1999).
The test particles were initially distributed smoothly in circular,
zero-inclination orbits between 28 AU and 63 AU, as indicated by the dotted
line in the lower panel.
In the upper two panels, the filled circles represent particles that remain
on stable orbits at the end of the simulation, while the open circles represent
the elements of those particles which had a close encounter with a planet
and subsequently move on scattered chaotic orbits (`removed' from the
simulation at the instant of encounter).  
Neptune's mean motion resonances are indicated at the top of the figure.}

\bigskip

\figcaption{
The temporal behavior of a long-lived member of the scattered disk.
The black curve shows the behavior of the comet's semi-major axis.
The gray curve shows the perihelion distance.
The three dotted curves show the location of the 3:13, 4:7, and 3:5
mean motion resonances with Neptune.}

\bigskip 

\figcaption{
The surface density of comets beyond Neptune for two different models
of the source of Jupiter-family comets.  The dotted curve is a model
assuming that the Kuiper belt is the current source (Levison \& Duncan~1997).
There are
$7 \times10^9$ comets in this distribution between 30 and $50\, AU$.  This
curve ends at $50\, AU$ because the models are unconstrained beyond
this point and not because it is believed that there are no comets
there.  The solid curve is DL97's model assuming the scattered disk is
the sole source of the Jupiter-family comets.
There are $6 \times10^8$ comets currently in this distribution.}

\bigskip 

\figcaption{
A strawman model of the Kuiper belt.  The dark area at left marked "Planets"
shows the distribution of solid material in the outer planets region:
it follows a power law with a slope of about -2 in heliocentric distance.
The dashed curve is an
extension into the Kuiper belt of the power law found for the outer
planets and illustrates the likely initial surface density distribution
of solid material in the Kuiper belt.  The dashed curve
has been scaled so that it is twice the surface density of the outer
planets, under the assumption that planet formation was 50\%
efficient.  The solid curve shows a model of the mass distribution by
Duncan et al$.$~(1995, see Figure 7), scaled to an estimated
population of $5 \times 10^9$ Kuiper belt objects between 30 and $50\, AU$.
The Kuiper belt is divided into three regions:, see text for a
description.  The dotted curves illustrate the unknown shape of the
surface density distribution in Region~B.}

\end{document}